\documentstyle[epsfig,12pt]{article}
\begin{document}
\begin{titlepage}
\begin{flushright}
ULB--TH 00/20\\
October 2000
\end{flushright}
\vspace*{1.6cm}

\begin{center}
{\Large\bf Neutrino focusing}\\
\vspace*{0.8cm}

D.~Monderen\\
\vspace*{0.2cm}

{\footnotesize\it 
Service de Physique Th\'eorique, Universit\'e Libre de
Bruxelles, CP 225, B-1050 Bruxelles, Belgium}\
\end{center}
\vspace*{1.0cm}

\begin{abstract}
We study the lensing of neutrinos by astrophysical objects. At the difference
of photons, neutrinos can cross a stellar core; as a result the lens quality improves.
While Uranians alone would benefit from this effect in the Sun, similar effects could be considered
for binary systems.
\end{abstract} 
\end{titlepage}

\section{Introduction}

Gravitational deflection of massless particles close to a massive object has
been extendedly studied in the case of photons and can be detected either
through multiple images or by signal enhancement \cite{weinberg,gl}.

Here due to the extreme difficulty to detect neutrinos and the comparatively
poor angular resolution of "neutrino telescopes", the only thing we
can hope for is a signal intensification.

While stars are notorously bad lenses for photons, we will show that the situation
can be very different for neutrinos: as they can cross even a stellar core, a real
focusing of a neutrino beam can be achieved.

\section{Neutrino deflection}

In ref.\cite{bigpaper}, we have computed the deflection for a neutrino beam crossing an
astrophysical object, as a function of the impact parameter $b$.

We simply give here the net deflection for a gaussian distribution cut at
the object's radius $R$ ($r_0$ is the gaussian width, $M$ the object's mass):

\begin{equation}
\label{gaussdefl}
\Delta\phi=
\left\{
\begin{array}{ll}
\frac{4M}{b} & {\rm if}\ b\geq R\\[2ex]
\frac{4M}{b}\left(1-\sqrt{1-\frac{b^2}{R^2}}\right)
+\frac{4M}{b}\frac{r_0/R\,e^{R^2/r_0^2}\sqrt\pi/2}
{r_0/R\,e^{R^2/r_0^2}\sqrt\pi/2\,{\rm erf}\left(R/r_0\right)-1}\\[2ex]
\times\left[\sqrt{1-\frac{b^2}{R^2}}\,{\rm erf}\left(R/r_0\right)-
e^{-b^2/r_0^2}{\rm erf}\left(\sqrt{1-\frac{b^2}{R^2}}\frac{R}{r_0}\right)
\right] & {\rm if}\ b<R
\end{array}
\right.
\end{equation}

A "good lens" requires that $\Delta \phi$ increases with $b$. This is more easily discussed
in terms of the effective focal length, $f(b)$.

\begin{figure}[ht]
\centerline{\epsffile{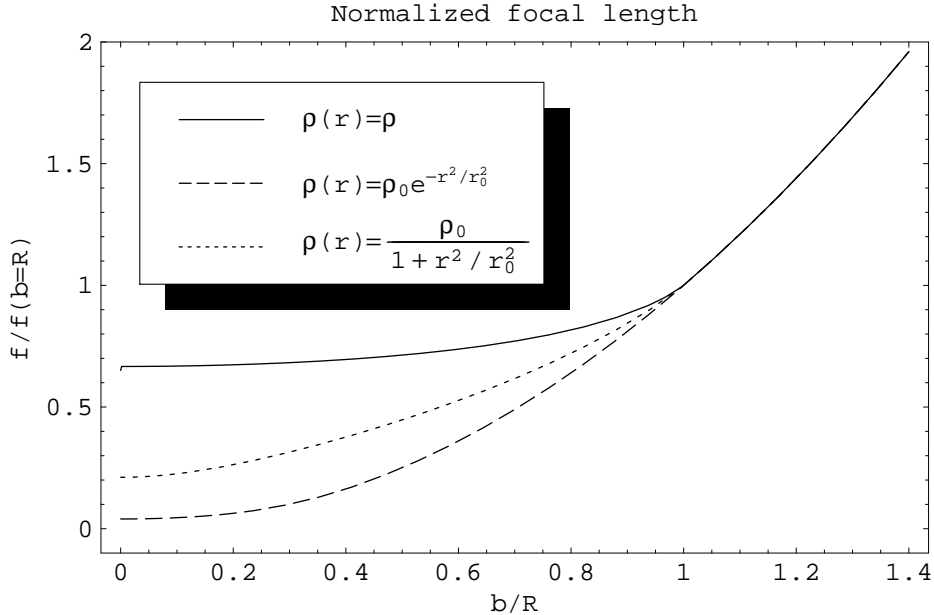}}
\caption{Normalized focal length $f/f(b=R)$ as a function 
of the normalized impact parameter $b/R$.} 
\label{focal} 
\end{figure}

Quite obviously, a perfect lens would correspond to a constant focal
length, whatever the distance from the center of the object. We give in Fig.~\ref{focal}
the (normalized) focal lengths $f(b)$ for various density profiles, namely a constant
density (typical for a planetary object), a gaussian distribution (which is an easily
tractable approximation of a stellar density profile), a Lorentzian repartition (evocative
of a galactic halo).

\section{Neutrino absorption}

While neutrinos are able to travel {\it across} the massive object, we have to
consider their interactions with the matter inside; these will indeed reduce
the neutrino flux and thus the efficiency of the signal amplification gained
through lensing. At the energies of interest for cosmic neutrinos ($E \ge 1-10$ GeV),
we have deep inelastic scattering of individual nucleons. Charged currents will cause
the conversion of neutrinos into charged leptons, which
will later be absorbed or decay, while neutral interactions will deflect them
by angles in general large compared to the lensing effect. In both cases, the interacting neutrinos
will be lost for lensing amplification.

Calling $\sigma$ the interaction cross-section (which depends on the neutrino
energy) and $N(x)$ the number density of scatterers along the neutrino path,
the neutrino flux attenuation after passing through a layer of matter of depth
$x$ is thus given by
\begin{equation}
\label{pro}
\Phi (x,E_{\nu}) = \Phi_0 e^{-\sigma (E_{\nu}) \int_0^x N(x')dx'}.
\end{equation}

Here we deal with objects having a spherical symmetry, so it is easier to express
Eq.~(\ref{pro}) in terms of the radial variable $r = \sqrt{x^2+b^2}$, $b$ being the impact
parameter of the neutrino with respect to the center of the object.
We then get
\begin{equation}
\label{proba}
\Phi (b,E_{\nu}) = \Phi_0 e^{-2\sigma (E_{\nu}) \int_b^R N(r)\frac{r
dr}{\sqrt{r^2 - b^2}}},
\end{equation}

\noindent where $R$ is the radius of the object. To solve this equation, we need to collect
information on the relevant cross sections, the possible mass density profiles
and the composition of the object (that is, how to relate mass density to number density).
For the cross sections, we refer the reader to Ref.~\cite{bigpaper}.

\bigskip

\begin{figure}[ht]
\centerline{\epsffile{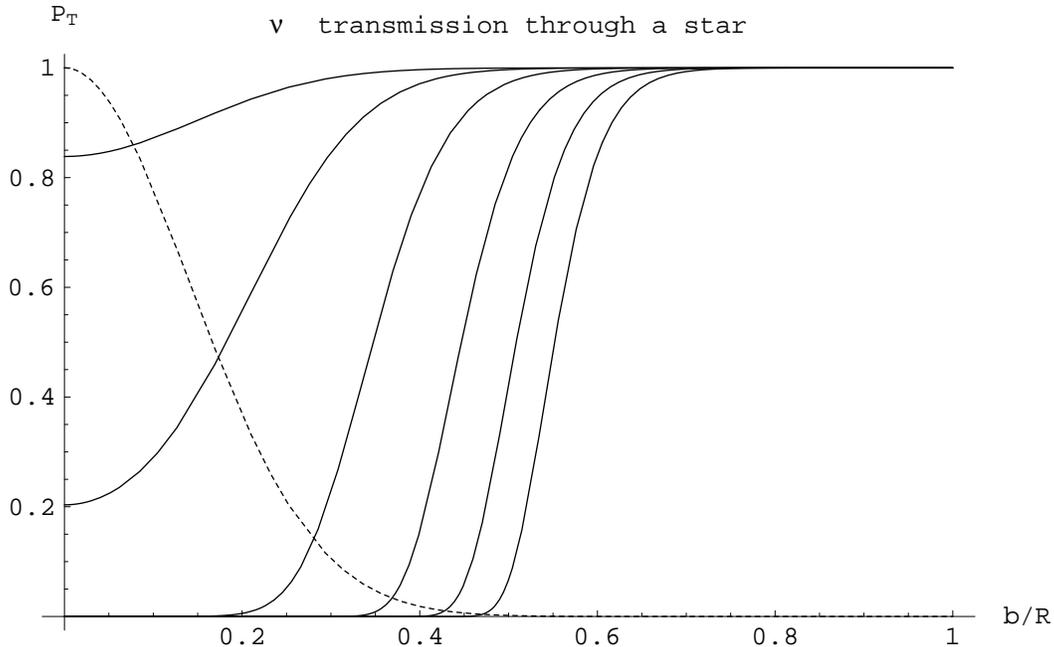}}
\caption{{Transmission probability in function of $b$, the impact parameter,
for a neutrino energy  $E_{\nu} =$ 10 GeV, $10^2$ GeV, $10^3$ GeV, $10^4$ GeV,
$10^5$ GeV, $10^6$ GeV  (solid lines, from left to
 right); the Gaussian
density profile $\rho(r) = e ^{-(\frac{5 r}{R})^2}$ is the dotted line.} }
\label{nu}
\end{figure}

We will suppose that the star\cite{ingthun} is made upof 25\% of Helium and 75\%
of Hydrogen in number; we then get the following relations between number
and mass densities:
\begin{eqnarray}
N_{He}(r) &=& 0.142 \, {\rm mol} \,\,\left( \frac{\rho(r)}{\rm 1 gr} \right)
\\ 
N_{H}(r)  &=& 0.427 \, {\rm mol}  \,\,\left( \frac{\rho(r)}{\rm 1 gr} \right)
\end{eqnarray}

\noindent that we used in our numerical calculations. We plot in Fig.~\ref{nu}
the results for neutrinos (a similar figure is obtained for antineutrinos).

\bigskip

Neutrinos passing through a galaxy may interact either with its visible matter or
with the surrounding halo of massive relic neutrinos. These last interactions
become significant only for incoming neutrinos at ultra high energies (of the order
of $10^{19}$ eV) \cite{roulet} and we may neglect them in our present frame of work.

Concerning the visible part of the Galaxy, a rough estimate gives an average density
of stars of 1 pc$^{-3}$, which corresponds to a negligible probability
for the neutrino to encounter a star during its passage through the galaxy,
even in the worst case if it traverses the whole disk and the bulge.

We thus conclude that the passage of neutrinos through the galaxy won't
decrease their flux, and hence do not affect the lensing effect.

\bigskip

We skip here the intricate but rather straightforward geometrical details, and jump to:

\section{Applications}

\subsection{The Sun}

We consider the Sun as a first example. Neutrinos passing outside the Sun
don't focus close enough to Earth to be of use; in fact for them to focus on Earth,
we would need

\begin{eqnarray}
b_0 \simeq 30 . 000 {\rm km} \simeq 5 \% R_\odot \, , 
\end{eqnarray}

\noindent which is clearly well inside the Sun, so the OUT solution does not provide
any sizeable effect.

We consider thus neutrinos crossing the Sun, using as announced a Gaussian
density profile for the matter distribution. As seen from Fig.~\ref{focal}, the plot
of the focal lenght never crosses the line $f=D_{LO}=1$ a.u. which would allow
for focusing on an earth-bound telescope. The smallest value is at about 22~a.u., which
means Uranians can perform wonderful neutrino lensing experiments using the Sun as lens,
see also \cite{gerver}. For them, any source would be amplified in turn as the Sun sweeps
in front of it! It is easy to check that Jupiter cannot replace the Sun as a useful lens for us,
as its mass is about $10^{-3} M_\odot$.

Even if we are not at the focal point, there is yet some amplification due to the improved
convergence of the beam. It provides

\begin{eqnarray}
\label{nofocus}
A = \left( \frac{f}{ \left| f-D_{LO} \right| } \right) ^2 \, .
\end{eqnarray}

\noindent The effect is significant if $D_{LO} \simeq f$, say $D_{LO} \in
\left[ 0.3 f ; 3 f \right]$, to have a magnification
higher than 2. Beyond $3f$, this last effect is negligible.

\subsection{Stars}

For distant sources, since the closest star is yet at about 1.3 pc, it is easy to check
(see above) that only neutrinos passing outside the star could be focused on Earth. The case
is similar to photons, namely the lens only focuses neutrinos passing through a thin ring
whose radius is fixed by the distance of the star. We have verified that for expected neutrino
sources the enhancement is insufficient to help in detection.

\subsection{Binary Systems}

For binary systems \cite{esteban}, if a compact source orbits close (ideally at the focal
distance) from a large star, the neutrino flux passing through this star can
be focused into a tight parallel beam (in a classical lighthouse picture).

Clearly, the result $f_{disk}/D_{LO} \approx 1$ is achievable. This situation is really
promising, providing the source is small enough compared to the lens, so that
the magnification is significant. Binary stars will however seldom meet all
the conditions; more exotic systems, with a compact and intense neutrino source, are needed.

\subsection{Galaxies}

In Ref.~\cite{bigpaper}, we have studied the lensing of both compact and galactic sources
by a galaxy. We conclude that in the former case, too few events can be observed, while the
second situation might be promising.

\section{Conclusions}

With an approach based on general relativity equations we have studied with
some generality neutrino lensing through different models of astrophysical
objects. The main difference with photons rests in the possibility for medium
energy neutrinos to cross even the stars (despite a cut-off above a few 100~GeV
in that case). This has the important result that the quality of the lens is greatly
improved, as in that case we get a nearly constant focal length in
the central region of the star. This would have been very promising, had the focal length
of the center of the Sun happened to coincide with the radius of the Earth orbit; unfortunately
this is far from being the case, and the focusing occurs closer to Uranus. For what concerns
the Lorentzian case, we notice that the result we obtained is significantly different
from the one usually quoted for photons passing through an isothermal potential.

We have considered other astrophysical objects than simple stars, and various
geometrical configurations, in order to identify the cases most favourable
for lensing. We have taken into account the finite size of the source and of the
detector and shown how they can influence the lensing characteristics.

The case of a binary system seems to be quite promising, provided small and energetic sources exist;
and galaxy lenses could also provide sizeable enhancements of the signal.

\section*{Acknowledgments}
This work was supported by I.~I.~S.~N. Belgium and by the Communaut\'e 
Fran\c caise de Belgique (Direction de la Recherche Scientifique programme 
ARC).
D.~Monderen benefits from a F.~R.~I.~A. grant.

\end{document}